\documentclass[pra,aps,preprint]{revtex4}
\usepackage[T1]{fontenc}
\usepackage[latin9]{inputenc}
\usepackage[active]{srcltx}
\usepackage{amsmath}
\usepackage{amssymb}
\usepackage{esint}
\usepackage{lineno}
\usepackage{color,soul}
\usepackage{xcolor}
\makeatletter
\usepackage{epsf,epsfig}
\usepackage{graphicx}

\makeatother
\begin{document}
\title{Laser induced plasma expansion in quantum plasma}
\author{Sepideh Dashtestani}
\author{Hamidreza Mohammadi} \thanks{hr.mohammadi@sci.ui.ac.ir} \affiliation{Department of Physics, University of Isfahan, Isfahan, Iran} \affiliation{Quantum Optics Group, University of Isfahan, Isfahan, Iran}
\date{\today}
\begin{abstract}
\noindent Interaction of ultera-short laser pulses with a dense cold plasma
is investigated. Due to high density, of plasma, quantum effects such that
Bohm potential and quantum pressure should be considered. The results reveal that electron density function
modulated by laser light in the propagation direction. This modulation can be controlled by
amplitude of laser intensity and plasma effective parameters.
For some special values of involved parameters electron density become localized in quenches
spatially. Increasing the quantum coefficient tends to rarefy the high electron density regions,
since the total number of electrons are constant. Hence, our theory predicts plasma expansion in the direction of laser light due to quantum effects.
\end{abstract}
\maketitle

\section{Introduction}
Recently, there has been a growing interest in investigating new aspects
of dense quantum plasmas \cite{Ref1}. Quantum plasmas have been achieved
in nano-scale objects such as nano-wires, quantum dots, quantum wells, and semiconductor
devices, as well as in laser plasma interaction \cite{Ref2} and wide interest due to
their applications to astrophysical and cosmological environments \cite{Ref3}. Usually,
the quantum plasma is characterized by a low-temperature and high-density plasma
\cite{Ref4}, when mean inter-particle distance is comparable with the deBorglie thermal wavelength of plasma.
Dense quantum plasma could be described by quantum hydrodynmics eqs.\cite{Ref5,Ref6} and quantum kinetic Eqs., including quantum forces
which are associated by Bohm potential \cite{Ref5} and quantum pressure
(due to degenerate electrons) which can be derived from Dirac's equation \cite{Ref7,Ref8,Ref9}.
On the other hand, Jung et.al show that when a laser pulse interacts with
a cold dense quantum plasma, new quantum feature appears \cite{Ref10}. This new quantum feature
is magnetization of plasma by photons which can be derived by considering a fiction 
time-dependent ponderomotive force.
Considering this new quantum aspect we obtain an electron density distribution
function for a cold isothermal dense quantum plasma illuminated
by an ultra-short intense laser pulse. The result reveal that density bunching effects occurs in high laser
intensities and for higher amount of quantum parameters. Our results predict plasma expansion due to quantum effects.
This paper is organized in four sections. In section 2, we have explained the models of interaction of laser pulse with
isothermal plasma.
The basic equations and fundamental assumptions are given for this model.
In section 3,
the profiles of the electric field and variation of electron density
are derived. Finally, section 4 is devoted to conclusion.

\section{Model description}
In this section we are going to analyze propagation of an intense
laser pulse through underdense collosionless quantum plasma in the
non-relativistic regime. The temperature of electrons
is held constant during the evolution. In this model, we consider a linearly 
x-polarized Gaussian laser pulse with time duration, $ \tau$, propagation
in +z direction in vacuum. The electron field of laser is written as:
\begin{equation} \label{eq1}
E(z,t)= \hat{x} E(z) e^{-i \omega t} =
\hat{x} E_0(z) e^{- \frac{t^2}{ \tau ^2} }  e^{-i \omega t}
\end{equation}
where $\omega$ is central angular frequency of the pulse. The $z > 0$ region is taken to be filled with a
homogeneous density profile of plasma with a plasma-vacuum
interface placed at z=0. The electromagnetic incident the plasma region
normally. Propagation of laser light in quantum plasma is governed
by Maxwell's equations:

\begin{equation} \label{eq2}
\overrightarrow{\nabla}  \times \overrightarrow{E}=- \frac{1}{c} \frac{\partial \overrightarrow{B}}{\partial t}, 
\end{equation}

\begin{equation} \label{eq3}
\overrightarrow{\nabla}  \times \overrightarrow{B}=- \frac{1}{c}  
\frac{\partial \overrightarrow{E}}{\partial t} + \frac{4\pi}{c} \overrightarrow{J},
\end{equation}

\begin{equation} \label{eq4}
\overrightarrow{\nabla}  . \overrightarrow{E}=4\pi \rho, 
\end{equation}

\begin{equation} \label{eq5}
\overrightarrow{\nabla}  . \overrightarrow{B}=0.
\end{equation}
Here, $\overrightarrow{J}=-n_ee\overrightarrow{\vartheta_e}$ is
the electron current density.
By taking into account Eqs. (2-5), and defining function $\Omega(t)=\omega +\frac{2t}{i \tau ^2}$,
for further simplicity we have:
\begin{equation} \label{eq6}
\frac{-\partial E_x}{\partial z} =
-i \frac{ (\Omega(t)+\dot{\Omega}(t)t)^2+2i \dot{\Omega}(t)}{c} B_y,
\end{equation}
\begin{equation} \label{eq7}
 \frac{-\partial B_y}{\partial z} =
-i \frac{ (\Omega(t)+\dot{\Omega}(t)t)^2+2i \dot{\Omega}(t)}{c} E_x- \frac{4\pi}{c} n_e e \vartheta _x.
\end{equation}
Where, $E_x$, $B_y$ and $ \vartheta _x$ are the component
of electric field in
x-direction, the component of magnetic field in y-direction and
electron velocity in x-direction.
Differentiating Eq. (6) with respect to z and
substituting it into Eq. (7), we obtain a reduced (Helmholtz)
differential equation for electric field in plasma:
\begin{equation} \label{eq8}
\frac{d^2}{dz^2} E_x+[ \frac{ (\Omega(t)+\dot{\Omega}(t)t)^2+2i \dot{\Omega}(t)}{c^2}]  \epsilon E_x=0.
\end{equation}
On the other hand the electron's equation of motion in
collisionless quantum plasma can be expressed as \cite{Ref11}:
\begin{equation} \label{eq9}
m_en_e[ \frac{\partial   \overrightarrow{\vartheta _e}}{\partial t} 
+(  \overrightarrow{\vartheta _e}.  \overrightarrow{\nabla} )  \overrightarrow{\vartheta _e}]=
-en_e \overrightarrow{E}-en_e \frac{  \overrightarrow{\vartheta _e} \times  \overrightarrow{B}}{c}
-n_ee  \overrightarrow{\nabla}  \phi -  \overrightarrow{\nabla} P_e-  \overrightarrow{\nabla} P_q 
\end{equation}
where, $m_e$ and $P_e=n_eT_e$
are electron mass and pressure of electron respectively.
$ \phi= \frac{eE_x^2}{m_e \omega ^2}$
is the average ponderomotive potential defined by the laser pulse envelope, with
its corresponding
ponderomotive force:
\begin{equation} \label{eq10}
 \overrightarrow{f_P}=-n_ee  \overrightarrow{\nabla}  \phi.
\end{equation}
Moreover, $P_q$ is quantum pressure including Bohm potential and pressure resulted from electron degeneracy.
Electron degeneracy pressure stems to Heisenberg's uncertainty relation.
Uncertainty of position and momentum implies that electron momentum is ill-defined and hence
electron continuesly move around its occupied position. This phenomenon exerts pressure on surrounding medium
which called electron degeneracy pressure.
According to \cite{Ref12,Ref13} we can write:
$$
 \overrightarrow{\nabla} P_q= \frac{ \hbar ^2}{2m_e} 
 \overrightarrow{\nabla} ( \frac{  \overrightarrow{\nabla ^2} \sqrt{n_e} }{ \sqrt{n_e} } )
- m  \overrightarrow{\vartheta _F^2}  \overrightarrow{\nabla} n^{(1)},
$$
where $n_e=n_e^{(0)}+n^{(1)}$, 
here $\overrightarrow{\vartheta _F}$ is Fermi's velocity.
In the following the motion in different direction can be written in the $1^{ST}$ order approximation as:
\begin{equation} \label{eq11}
m_e( \frac{\partial  \vartheta _x}{\partial t} )=-eE_x,
\end{equation}
\begin{equation} \label{eq12}
- \frac{eB_0 \vartheta _x}{c}- \frac{e\partial \phi }{\partial z} 
- \frac{1}{n_e }\frac{dP_e}{dz}  - \frac{1}{n_e} \frac{dP_q}{dz}=0.
\end{equation}
Where $B_0$ and $n_{0e}$ are refereed to external magnetic field (which
is considered in the direction of laser field) and the maximum electron density respectively.
By using Eqs. (8-12) we have:
\begin{equation} \label{eq13}
 \frac{\partial}{\partial z}[ \frac{-e^2E^2_x-B_0B_y}{2m_e^2 (\Omega(t))^2\Omega _q}]
= \frac{\overrightarrow{ \nabla }n_e}{n_e}  
\end{equation}
And hence:
\begin{equation} \label{eq14} 
n_e=n_{0e}exp[ \frac{-e^2(E^2_x-B_0B_y)}{2m_e^2 (\Omega(t))^2  \Omega _q} ].
\end{equation}
This equation shows that the electron density distribution
depends on the laser electric field
$E^2_x$, laser magnetic field $B_y$ and external magnetic field
$B_0$.
When the electric and magnetic fields are zero in the Eq. (14), $n_e$
represents the maximum electron density as we expected. Also we defined $\Omega_q$ as:
$$ \Omega _q=\frac{T_e}{m_e} +   \overrightarrow{\vartheta _F^{2}} + \frac{ \hbar ^2K^2}{4m_e^2}. $$
The quantum mechanical effects are encapsulated in the last two terms of this equation.
In the classical limit ($\hbar \rightarrow 0)$ the results of non-quantum plasma is achieved \cite{Ref14}.
On the other hand,
non-magnetized plasma
$(B_0=0)$, Eq. (14) turned to:
\begin{equation} \label{eq15}
n_e=n_{0e}exp[ \frac{-e^2E^2_x}{2m_e^2 (\Omega(t))^2  \Omega _q} ]
\end{equation}
The results is in agreement with the results of ref \cite{Ref15} for non-magnetized case
in the absence of quantum effects.
In literature of this field, it is convent to start from
$P_e=n_eT_e$ and $ \overrightarrow{f_P}=  \overrightarrow{\nabla} P_e$, to extract the electron density $n_e$
\cite{Ref16,Ref17}. Reversing this procedure and starting from $n_e$ from
Eq. (15), and $P_{total} = P_e+P_q$ we can write:
\begin{equation} \label{eq16}
 \overrightarrow{f_P}=  \overrightarrow{\nabla} P_e+ \overrightarrow{\nabla} P_q
\end{equation}
for quantum plasma.
Is this Eq. (16) includes all important quantum effects?
The answer is actually No!
For example, the quantum mechanic predict an induced magnetic field due to
considering quantum feature of plasma dielectric constant. Such induced magnetic fields
are introduced in \cite{Ref10}.
In this paper, firstly we show that this purely quantum mechanical induced
magnetic field can be derived by considering a fiction time-dependent pondromotive force.
In the following we try to associate a electron density function to the plasma
which describes the interaction of laser with quantum plasma.
\begin{equation} \label{eq17}
f_P^{(t)}=  \frac{-e^2E^2_x}{2m_e^2 (\Omega(t)) ^2 \Omega _q}
\frac{2k \omega_p^2}{(\Omega(t)) ^3}
 \frac{\partial  | E | ^2}{\partial t}
\end{equation}
This ponderomotive force relates to the electron density $n_e$ via $\omega_p$,
this fact increases complexity of calculation. In order to overcome to this problem
we replace $\omega_p$ by $\omega_{p0}$ in the last equation.
Considering this new type of ponderomotive force modifies Eqs. (13) and (16) as follow:
\begin{equation} \label{eq18}
 \frac{\partial}{\partial z}[ \frac{-e^2E^2_x}{2m_e^2 (\Omega(t))^2\Omega _q} ]
+ \frac{\partial}{\partial t} 
 [\frac{-e^2E^2_x}{2m_e^2 (\Omega(t))^2 \Omega _q}
\frac{2k \omega_p^2}{(\Omega(t))^3}| E | ^2]
= \frac{\overrightarrow{ \nabla }n_e}{n_e},  
\end{equation}
\begin{equation} \label{eq19}
\overrightarrow{f}_P^{(s)}+ \overrightarrow{f}_P^{(t)}=
\overrightarrow{\nabla} P_{total}= \overrightarrow{\nabla} P_e+ \overrightarrow{\nabla} P_q.
\end{equation}
For unmagnetized quantum plasma ($B_0=0$), and for Gaussian
laser pulse this equation leads to the following formula:
\begin{equation} \label{eq20}
\frac{-e^2E^2_x}{2m_e^2 (\Omega(t))^2\Omega _q} 
+\frac{2ie^2k_z \frac{\omega_p ^2}{(\Omega(t))^2}}{2m_e^2 (\Omega(t))^2\Omega _q}  \int_0^L  E^2_x| E | ^2dz
=ln \frac{n_e}{n_{0e}},
\end{equation}
\begin{equation} \label{eq21}
n_e=n_{0e}\hspace{.02cm}exp(\frac{-a^2}{2} 
+2ib^2k_z \frac{c}{\Omega(t)}\int_0^{L\frac{\Omega(t)}{c}}  \frac{-a^2}{2} | E | ^2d\xi).
\end{equation}
Where
$$a^2= \frac{e^2E_x^2}{m_e^2 (\Omega(t))^2 \Omega _q}, \hspace{4mm}
 \xi = \frac{z \Omega(t)}{ c },\hspace{4mm} and \hspace{4mm}
b= \frac{\omega_p}{\Omega(t)}$$
are dimensionless variables.
Starting from Eq. (1) for a exponentially descending field i.e.
$E(z)=E_0e^{-\alpha z}( \alpha  > 0)$, ($\alpha$ is absorption coefficient of the
medium with the dimension of $L^{-1}$), we have
$\overrightarrow{E} = \hat{x} E_0 e^{-i \beta Z}e^{-i \Omega(t) t}$, where
$\beta=-i\alpha+k_z$ is complex-valued wave number. Hence,
$$ | E | ^2=exp(- \frac{2c \beta \xi}{ \Omega(t) } )I(0)$$
\begin{equation} \label{eq22}
n_e=n_{0e}exp(\frac{-a^2}{2} 
+2ik_z\frac{\omega_p^2}{(\Omega(t))^2}I(0)
\frac{c}{\Omega(t)}\int_0^{L\frac{\Omega(t)}{c}}  \frac{-a^2}{2} 
exp(- \frac{2c \beta \xi}{ \Omega(t) } )d\xi).
\end{equation}
By considering Eq. (8) and (22) dielectric constant of quantum plasma
is derived as:
\begin{equation} \label{eq23}
 \varepsilon =1- \frac{ \omega _{p0e}^2}{ (\Omega(t))^2} exp(\frac{-a^2}{2} 
+2ik_z\frac{\omega_p^2}{(\Omega(t))^2}I(0)
\frac{c}{\Omega(t)}\int_0^{L\frac{\Omega(t)}{c}}  \frac{-a^2}{2} 
exp(- \frac{2c \beta \xi}{ \Omega(t) } )d\xi),
\end{equation}
and ultimately Eqs. (8) and (22) yield governing differential equation of
propagation as:
$$\frac{d^2a}{d \xi ^2} +\frac{(\Omega(t)+\dot{\Omega}(t) t)^2+2i\dot{\Omega}(t)}{(\Omega(t))^2}\times$$
\begin{equation} \label{eq24}
[1- \frac{ \omega _{p0e}^2}{(\Omega(t))^2}
 exp(\frac{-a^2}{2} 
+2ik_z\frac{\omega_p^2}{(\Omega(t))^2}I(0)
\frac{c}{\Omega(t)}
\int_0^{L\frac{\Omega(t)}{c}}  \frac{-a^2}{2} 
exp(- \frac{2c \beta \xi}{ \Omega(t) } )d\xi)]a=0.
\end{equation}
Starting from Eq. (22) and by some algebra we can write:
\begin{equation} \label{eq25}
n_e=n_{0e}exp[\frac{-a^2}{2} 
-(1+i(1+\pi)) | W |  exp(-2i \Omega(t) t) exp(i\pi) exp(-2|\beta| L(1+i)-1)]  
\end{equation}
Where $W=\frac{e^2 E_0^2}{4 m_e^2 c \Omega_q} \frac{bd}{\beta}$ with $d=\frac{K_z I(0) c}{\Omega(t)}$.
The quantum effect is encapsulated in W so we call W as quantum factor.
\begin{figure}[!htb]
\includegraphics[width=10cm]{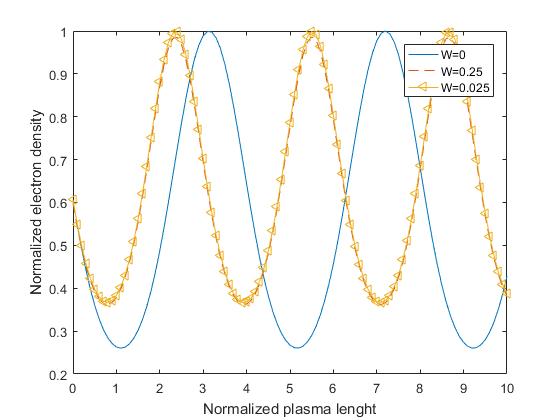}
\caption{Vriations of normalized electrons density $\frac{n_e}{ne0}$ as a function of normalized
plasma length $\frac{\Omega z}{c}$ inside isothermal plasma for different quantum coefficients W=0 (solid line), W=0.25 (dotted line), and W=0.025 (dashed line) for $a_0=1, b=0.1$.}
  \label{1}
\end{figure}
\begin{figure}[!htb]
\includegraphics[width=10cm]{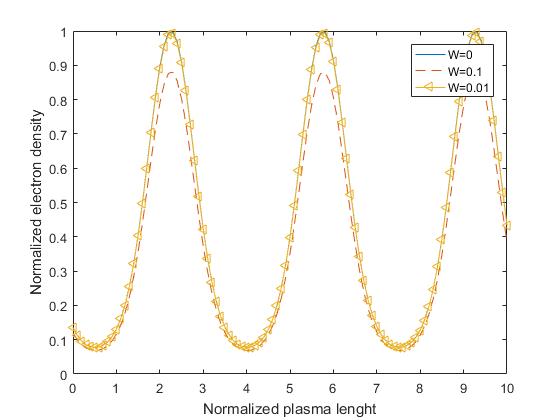}
\caption{Vriations of normalized electrons density $\frac{n_e}{ne0}$ as a function of normalized
plasma length $\frac{\Omega z}{c}$ inside isothermal plasma for different quantum coefficients W=0 (solid line), W=0.1 (dotted line), and W=0.01 (dashed line) for $a_0=2, b=0.1$.}
  \label{2}
\end{figure}
\begin{figure}[!htb]
\includegraphics[width=10cm]{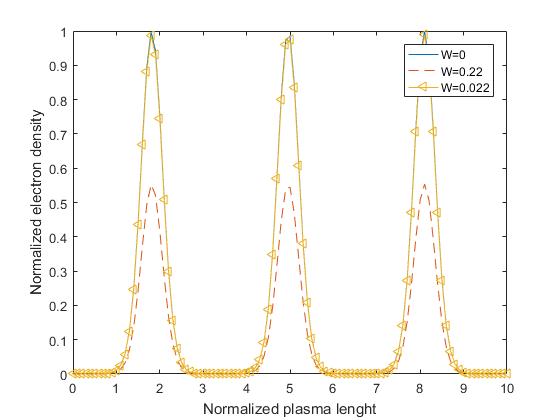}
\caption{Vriations of normalized electrons density $\frac{n_e}{ne0}$ as a function of normalized
plasma length $\frac{\Omega z}{c}$ inside isothermal plasma for different quantum coefficients W=0 (solid line), W=0.22 (dotted line), and W=0.022 (dashed line) for $a_0=3, b=0.1$.}
  \label{3}
\end{figure}
\begin{figure}[!htb]
\includegraphics[width=10cm]{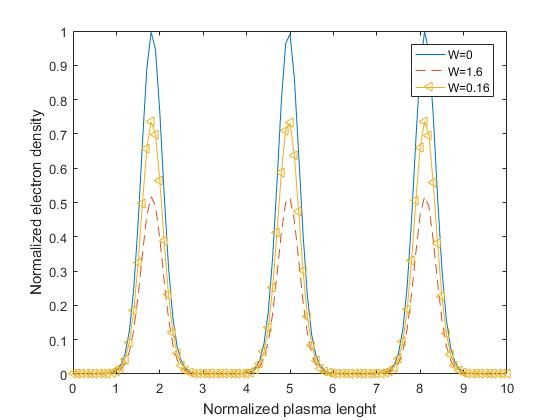}
\caption{Vriations of normalized electrons density $\frac{n_e}{ne0}$ as a function of normalized
plasma length $\frac{\Omega z}{c}$ inside isothermal plasma for different quantum coefficients W=0 (solid line), W=1.6 (dotted line), and W=0.16 (dashed line) for $a_0=4, b=0.4$.}
  \label{4}
\end{figure}

\section{Results}
Due to nonlinearity of governing Eq. (25), analytical trace of calculation is impossible.
The results of numerical solution of Eq. (25) is illustrated in Fig. (1-4).
In these figures we focus on the variation of electron density through the plasma. Figure are plotted
for some values of laser intensity amplitude, size of quantum factor and b=$\frac{\omega_p}{\Omega(t)}$.
These parameters are not independent (see definition of W), so changing one parameters affect the value of others.
All of the figures, show that electrons accumulate in some region and hence some regions deplete from
free electrons. The loci of these regions and their electron number can be altered by introducing
quantum effects. This phenomenon intensively affects the transport process of the plasma
(such as electrical conductivity and thermal conductivity).
Figure (1) depicts that electron density function oscillates through the plasma
in the direction of laser light. Quantum effects decreases the wavelength of this oscillation.
Variation of normalized electron density versus normalized plasma length is illustrated in
figure (2) for intermediate laser intensity.
In this regime departure from classical plasma is not considerable.
Increasing the intensity of laser light reveal quantum effects.
Figures (3) and (4) show this fact.
Further increasing of W, cause to decrease peak of $n_e$. Since the total electron
number of plasma is constant, decreasing in $n_e$ is accombined by increasing the volume (length)
of plasma.
Also, for high intensities of laser light, similar to classical
plasma $n_e$ completely localized in some regions in the direction of laser light.
This fact my be is due to solitary solution of non-linear Eq. (24).

\section{Conclusion}
A new phenomenon is observed in the process of laser-plasma  interaction.
We claim that when an ultra-short Gaussian laser pulse interact with a quantum plasma.
The size of plasma in the laser direction is changed. This phenomenon has various 
applications. For example it can be used in performance of Acousto-optical modulators (AOM).

\end{document}